# Magnetoelectric Control of Toroidal Moment in Ferroaxial Crystal PbMn$_2$Ni$_6$Te$_3$O$_{18}$

Shungo Aoyagi[1,2], Shunsuke Kitou[3], Taka-hisa Arima[3], and Yusuke Tokunaga[3,*]

[1] *Department of Applied Physics, University of Tokyo, Tokyo 113-8656, Japan*

[2] *Institute of Industrial Science, University of Tokyo, Tokyo 153-8505, Japan*

[3] *Department of Advanced Materials Science, University of Tokyo, Kashiwa, Chiba 277-8561, Japan*

*E-mail: y-tokunaga@edu.k.u-tokyo.ac.jp

**Abstract**

Ferroic multipole orders break some symmetry and often activate unique physical properties, such as nonreciprocal directional dichroism and linear magnetoelectric effects. Here, we demonstrate the control of magnetic toroidal orientation in PbMn$_2$Ni$_6$Te$_3$O$_{18}$ through the application of electric and magnetic fields in two distinct configurations. Through directional dichroism, we successfully visualize magnetic ferrotoroidic domains, establishing the intrinsic coupling among magnetic toroidal moment, crystallographic ferroaxial moment, and magnetoelectric monopoles. Our findings not only present an effective pathway for controlling magnetic toroidal moment but also provide a novel approach for investigating ferroaxial ordering.

Cross-correlation phenomena in condensed matter, such as magnetoelectric (ME) effect and thermoelectric effect, are fundamental to the development of multifunctional materials for electronic devices. Recent advances in understanding these phenomena in the multipole framework[1–4] are useful for designing various material responses, including Seebeck effect[5,6], Nernst effect, and certain types of superconductivity[7,8].

Ferroic multipole orders have inherent bistability with the sign reversal of the order parameter, showing promise for the applications to information devices such as magnetoresistive random access memory (MRAM). From the perspective of device application, control and observation of the sign of multipole are essential. Recently, developing domain control methods utilize non-conjugate external fields through intrinsic free-energy couplings[9], and target high-order multipoles such as altermagnets[10]. Similarly, domain observation techniques have diversified to apply to high-order multipole, such as ferroaxial order[11–14].

The linear ME effect, expressed as **P**=α**H** where **P** is the electric polarization and **H** is the magnetic field, represents a well-established manifestation of multipole physics. The origin of ME tensor $\alpha$ can be decomposed into three kinds of multipole moments: ME monopole[15], magnetic toroidal dipole, and magnetic quadrupole[16]. Notably, optical ME effects such as directional dichroism, a change in the absorption coefficient of unpolarized light with the reversal of propagating direction, have been demonstrated to facilitate a convenient detection of magnetic toroidal domains[17]. While this technique offers a straightforward approach to domain detection, its potential for indirect domain observation in more complex systems has remained underexplored.

In this study, we demonstrate another domain engineering approach to controlling magnetic toroidal order through the coupling of multiple multipole orders in hexagonal PbMn$_2$Ni$_6$Te$_3$O$_{18}$ (Fig. 1(a)), which exhibits collinear antiferromagnetic order parallel to the $c$-axis below $T_N$=84 K[18–22], as shown in Fig. 1(b). Below $T_N$, it belongs to the magnetic point group 6/$m'$, which is not ferroelectric but linear magnetoelectric[18] and allows the coexistence of ferroic orders of ME monopole ($M_0$) and magnetic toroidal dipole along the $c$-axis ($T_z$)[21]. Since the magnetic moments of Mn sites ($S$=5/2) and adjacent Ni sites ($S$=1) are ferromagnetically coupled, only two types of antiferromagnetic domains exist as reported by Doi *et al.*[19]. Because of the unique crystal structure that pairs of NiO$_6$ clusters exhibit opposite $c$- axis component of electric dipole moment (Fig. 1(c)) and chirality (Figs. 1(d) and (e)), ferroaxial moment parallel to the $c$-axis ($A_z$) exists as in-plane components cancel among sites that are related by the threefold rotation. In addition, Mn and Ni sites related by a mirror symmetry operations possess opposite magnetic moments along the $c$-axis. This crystal therefore creates inherent couplings among ferroaxial order, magnetic toroidal order, and ME monopole order, providing an ideal platform for investigating multipole domain engineering. In general, this coupling can emerge in ferroaxial crystals that possess a single rotational axis and a mirror plane perpendicular to this rotational axis as symmetry elements. This occurs when the system exhibits a collinear antiferromagnetic order with spins oriented parallel to the rotational axis and two magnetic sites linked by a mirror symmetry have magnetic moments pointing in the opposite directions.

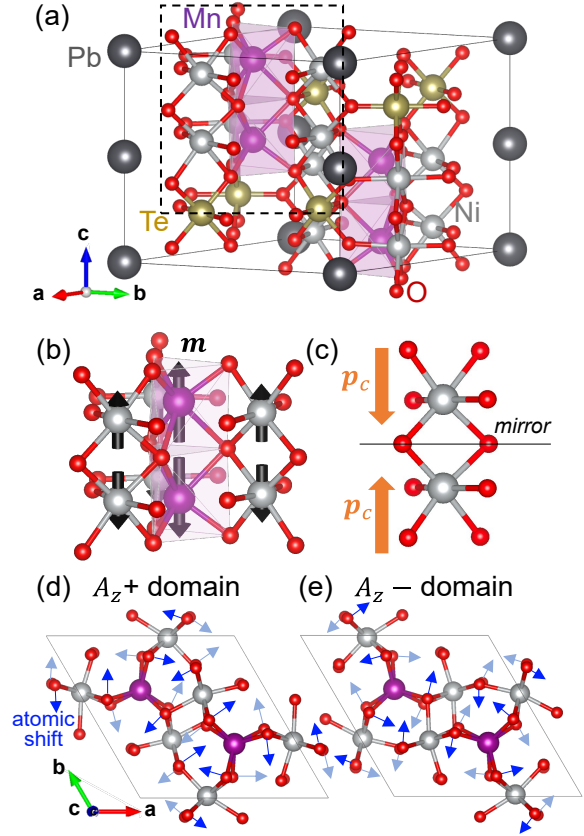

**FIG. 1.** (a) Crystal structure of $PbMn_2Ni_6Te_3O_{18}$[19]. (b) Magnetic structure of $PbMn_2Ni_6Te_3O_{18}$[19] below $T_N = 84$ K in the part enclosed by the dashed rectangle in (a). The magnetic moment is denoted by ***m***. (c) Two $NiO_6$ octahedra related by a mirror operation. Orange arrows represent the $c$-axis component ***p***$_c$ of electric dipole induced by the displacement of Ni site. (d) $c$-axis projection of the part enclosed by the dashed rectangle in (a). (e) A mirror image of (d) about a plane $x = y$. The breaking of the mirror reflection may allow the presence of ferroaxial domains $A_z^+$ and $A_z^-$. Blue (light blue) arrows indicate the displacement along the $c$-plane of oxygen atoms at upper (lower) part of each cluster, which produces local chiralities. All figures are drawn using VESTA[23].

Employing this coupled-order approach, our experiments demonstrate that magnetic toroidal domains can be effectively manipulated through various field arrangements. The intrinsic coupling among ferroaxial domains, magnetic toroidal domains, and ME monopoles is also confirmed. By successfully visualizing magnetic toroidal domains using directional dichroism, we provide a powerful method for investigating ferroaxial ordering and multipole coupling phenomena that could advance the development of novel electronic devices.

Single crystals of PbMn$_2$Ni$_6$Te$_3$O$_{18}$ were synthesized using the self-flux method. First, powders of PbO, MnCO$_3$, NiO, and TeO$_2$ were mixed in a molar ratio of 1:3:5:5 and placed in an alumina crucible. The mixture was heated at 650 °C for 12 hours, then at 850 °C for 24 hours, followed by cooling at a rate of 1.5 °C/h over 100 hours. Subsequently, the obtained polycrystals were mixed with TeO$_2$ in a molar ratio of 1:4 and reheated following the same procedure, resulting in single crystals of approximately 1 mm on each side (see supplementary material Fig. S2).

Electric polarization was measured with an electrometer (Keyithley 6517A) in a cryostat equipped with a superconducting magnet (Spectromag, Oxford). The single crystals were cut into plates with large $c$ planes or $(1\bar{1}0)$ planes, thinned to 0.1 mm, and polished by Al$_2$O$_3$-coated lapping films. Magnetic field dependence of polarization was measured at 80 K, below $T_N$, with a magnetic field of $\mu_0 H = 8$ T and an electric field of $E_{pol} = 1.8$ MV/m, both applied during the cooling process. During the measurement, the electric field was removed, and the magnetic field was swept at a rate of 1 T/min. Measurements of quasi-static ferroelectric hysteresis loops ($P$-$E$ loops) conducted also at 80 K with a magnetic field of $\mu_0 H = 8$ T. During cooling to 80 K, no external field was applied.

For the optical absorption measurement, the single crystals were cut into plates with large $c$ planes, thinned to 0.08-0.15 mm, and polished by Al$_2$O$_3$-coated lapping films. Monochromatized light from a halogen lamp was propagated along the $c$-axis of the sample. The transmitted light intensity was detected by a photodiode for spectroscopy and a charge-coupled detector camera (Alta U6, Apogee Instruments) for imaging. Electric and magnetic fields were applied using a high-voltage supply (Model 2657A, Keithley) and an electromagnet, respectively. Poling was performed using external fields in two distinct configurations during cooling the sample from 150 K. In the first configuration, a poling electric field of $E_{pol}=0.58$ MV/m was applied perpendicular to the $c$-axis through 100 Å thick gold electrodes deposited on the upper and lower surfaces of the optical path. Simultaneously, a magnetic field of $\mu_0 H_{pol}=0.1$ T was applied perpendicular to both the optical path and electric field. In the second configuration, both electric ($E_{pol}=1.8$ MV/m) and magnetic ($\mu_0 H_{pol}=0.1$ T) fields were applied parallel to the $c$-axis. The electric field was applied through 60 Å thick gold electrodes deposited across the entire $c$-plane surface. During the measurement, both the electric field and magnetic field were removed.

First, we evaluated the ME properties of PbMn$_2$Ni$_6$Te$_3$O$_{18}$. Figure 2(a) shows the magnetic field dependence of electric polarization for three components. These results confirm the presence of linear ME effect as predicted. The linear ME tensor allowed for the magnetic point group 6/$m$' is:

$$\alpha = \begin{pmatrix} \alpha_\perp & \alpha_t & 0 \\ -\alpha_t & \alpha_\perp & 0 \\ 0 & 0 & \alpha_\parallel \end{pmatrix}.$$

The largest ME coefficient is observed for $\alpha_\parallel$, followed by $\alpha_t$ and $\alpha_\perp$ in descending order.

Next, we study $P$-$E$ loops in an applied magnetic field. Figure 2(b) shows a $P$-$E$ loop for the $\alpha_\parallel$ component, starting from a zero-field-cooled state. Based on the relationship between multipoles and the ME tensor, this electric polarization can be attributed to the ME monopole and the magnetic quadrupole whose contributions are clear from the distinct values of $\alpha_\parallel$ and $\alpha_\perp$ (for details, see supplementary materials). In this material, the quadrupole phase (0 or $\pi$) has a one-to-one correspondence with the ME monopole phase; therefore, for simplicity, we will ignore this in the following. The gray dots indicate the initial electric field sweep process, showing the transition from a multi-domain state to a single-domain state with increasing the electric field. The blue dots depict a typical P-E hysteresis loop, indicating that the ME monopole domains can be reversed by sweeping electric field in a magnetic field. Figure 2(c) shows a P-E loop related to $\alpha_t$, which arises from the magnetic toroidal dipole. This loop exhibits a shape similar to Fig. 2(b), demonstrating that the magnetic toroidal dipole domains can also be reversed by electric field sweeping. Furthermore, the magnitudes of these polarizations are consistent with the results shown in Fig. 2(a). Thus, we successfully observed the reversal of polarization purely arising from magnetic toroidal moments without quadrupole contributions. While toroidal component reversals have been observed in materials such as LiCoPO$_4$[24,25], this represents the first demonstration of purely toroidal-derived polarization reversal without any quadrupole contributions.

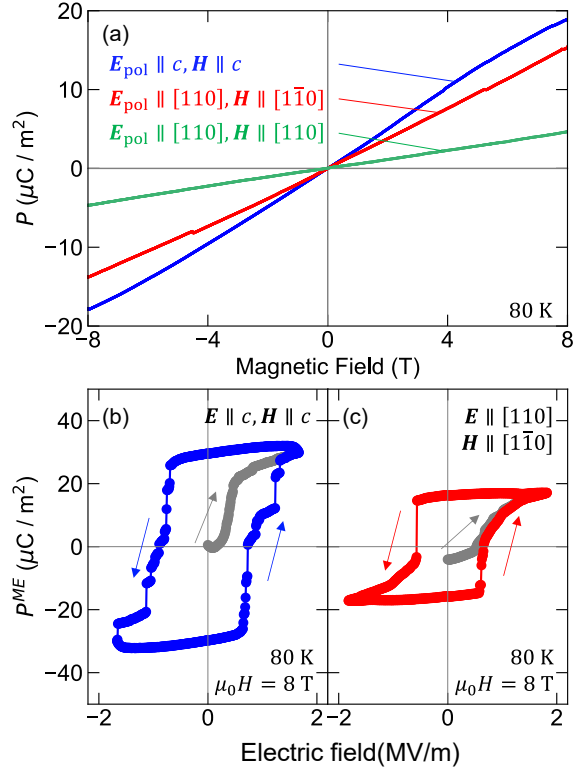

**FIG. 2.** (a) Magnetic field dependence of electric polarization at 80 K ($<T_N$) after cooling in poling electric ($E_{pol}$) and magnetic fields ($H_{pol}$). $E_{pol}$ (∥$P$) =1.8 MV/m and $\mu_0 H_{pol}$ (∥$H$) = 8 T was applied during cooling. Time-proportional offset due to constant current flow during the magnetic field sweep process has been removed. (b), (c) Electric field dependence of polarization when sweeping the electric field in an 8 T magnetic field, starting from a zero-field-cooled state, with external fields applied in the directions (b) $E \parallel H \parallel [001]$ and (c) $E \parallel [110]$, $H \parallel [1\bar{1}0]$. The gray curves represent the change in polarization when gradually applying the electric field from the initial state. Here, the component of electric polarization proportional to the electric field has been removed.

We proceed to visualize magnetic toroidal domains to clarify the coupling between these multipole orders. Figure 3(a) shows the measurement setup. Figure 3(b) shows the absorption coefficient spectra at 80 K and 110 K for the configuration shown in Fig. 3(a), where the magnetic toroidal dipole is aligned parallel to the light propagation. A large absorption peak is observed around 1.7 eV, which is attributed to a d-d transition of Ni$^{2+}$ [22]. Figure 3(c) displays the difference in absorption spectra between two configurations with opposite directions of magnetic toroidal dipoles. In the present configuration, the electric and magnetic fields of the light lie within the $ab$ plane, so that the off-diagonal components $\alpha_{xy}$ of the ME tensor affect the absorption coefficient. Under the symmetry of the magnetic point group $6/m'$, these off-diagonal components originate solely from the magnetic toroidal dipole[26]. While no difference is observed in the paramagnetic phase (110 K), a peak appears around 1.58 eV in the antiferromagnetic phase (80 K). We note that the difference between the two spectra in Fig. 3(b) includes contributions from the temperature change as well as from the magnetic ordering, and therefore does not necessarily correspond to the spectral difference shown in Fig. 3(c). This result indicates the presence of directional dichroism, as the absorption spectrum depends on whether the light propagates parallel or antiparallel to the magnetic toroidal dipole. Here, the magnetic moments on the Ni sites are along the light propagation. Considering the local chiralities of Ni sites, this can be understood as magnetochiral dichroism (MChD)[27–29]. MChD is an optical effect microscopically arising from the interference between electric dipole and magnetic dipole transitions. The dichroism observed here is notably larger than that of $MnTiO_3$[27,29], with a similar symmetry. Notably, from a symmetrical perspective, directional dichroism is not derived from the ME monopole or magnetic quadrupole but magnetic toroidal dipole in this configuration. Measurements conducted in setups where external fields are applied parallel to the $c$-axis make the ME monopole single domain state, but not always align magnetic toroidal dipole, as discussed later.

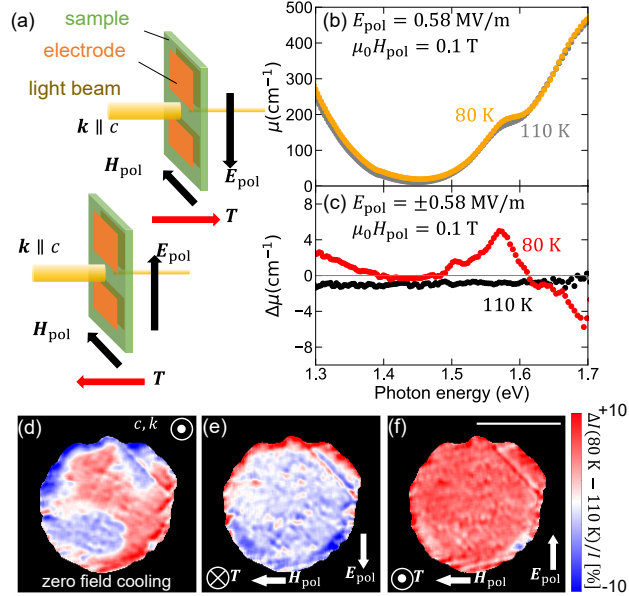

**FIG. 3.** (a) Setup for directional dichroism measurement. $E_{pol} = \pm 0.58$ MV/m, and $\mu_0 H_{pol} = 0.1$ T. The direction of the induced magnetic toroidal dipole is reversed by changing the direction of the applied electric field. Measurements are taken after removing the external fields. (b) Spectra of absorption coefficient at 80 K (antiferromagnetic state) and 110 K (paramagnetic state). Here, $d = 110$ μm. (c) Difference in spectra of absorption coefficient between the two poling conditions shown in (a). (d)-(f) Differential images between 80 K and 110 K under several poling field conditions: (d) $E_{pol}=H_{pol}=0$, (e) $(E_{pol}\times H_{pol})\cdot k<0$ and (f) $(E_{pol}\times H_{pol})\cdot k>0$. The energy of incident light is 1.58 eV.

Next, we perform imaging using light of a photon energy 1.58 eV. Figure 3(d) shows the differential image between 80 K and 110 K when the sample undergoes the transition at zero external electric and magnetic fields. The method for calculating the differential image is described in the supplementary materials. The image is composed of two types of domains: red region of weak absorption, and blue region of strong absorption. The difference in brightness is consistent with the results in Fig. 3(c). These results demonstrate the existence of magnetic toroidal domains of approximately 100 μm.

Figure 3(e) shows the result obtained when the antiferromagnetic transition takes place under poling electric and magnetic fields. The whole region exhibits the same blue coloration, indicating the magnetic-toroidal single domain state. Additionally, as shown in Fig. 3(f), the poling electric field in the opposite direction results in the reversal of the signal corresponding to the reversal of the induced magnetic toroidal dipoles. This result allows us to clarify the relationship between the sign of the magnetic toroidal moment and the change in absorption.

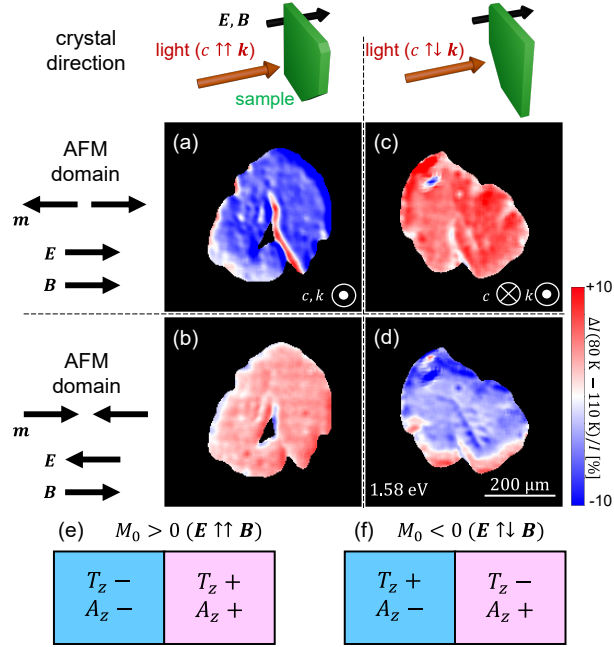

**FIG. 4.** (a)-(d) Temperature differential images under four conditions with two types of antiferromagnetic domain orientations and two types of crystal orientations. As shown in the top, external fields and light propagation are parallel or antiparallel to the *c*-axis. The photon energy of incident light is 1.58 eV. In (a) and (b) [(c) and (d)], the antiferromagnetic Néel vector is aligned in opposite directions by flipping the direction of the poling electric field. Between (a) and (c) [(b) and (d)], the crystal *c*-axis is inverted by a 180 degrees rotation of the crystal. The color change at the sample edges is an artifact due to a slight shift of the sample by cooling. (e), (f) Schematic drawing of ferrotoroidic and ferroaxial domains with aligned ME monopole domain.

Next, we study the effect of poling electric and magnetic fields parallel to the *c*-axis, which can align ME monopole. Figure 4(a) shows the differential image between 80 K and 110 K. Furthermore, as shown in Fig. 4(b), when we reverse the direction of the poling electric field to create oppositely oriented ME monopole single-domain state, we observe that the magnetic toroidal is also reversed. This experiment demonstrates the complete coupling between the ME monopole and magnetic toroidal moments.

We then conduct a similar experiment after flipping the *c*-axis by rotating the sample by 180 degrees around an axis perpendicular to the *c*-axis. This two-fold rotation is equivalent to a mirror operation with respect to a plane containing the *c*-axis from a symmetry perspective. A two-fold rotation about the axis perpendicular to the *c*-axis can be written as, for example, $(x, y, z) \to (-x, y, -z)$. Applying the mirror operation with respect to a plane perpendicular to the *c*-axis, which is a symmetry operation of the point group 6/*m*, this becomes $(-x, y, -z) \to (-x, y, z)$, entirely equivalent to a mirror operation with respect the *bc* plane, (*x*, *y*, *z*) → (-*x*, *y*, *z*). This operation therefore corresponds to a reversal of the ferroaxial domain. Microscopically, under the two-fold rotation, the arrangement of the magnetic moments of Mn and Ni shown in Fig. 1(b) remains invariant, whereas the chiral arrangement of the MO$_6$ clusters is reversed.

While a natural question may arise as to whether a toroidal moment can emerge from a collinear magnetic order, by applying this domain flipping, we indeed observe a reversal of the directional dichroism coupled with the magnetic toroidal moment, as shown in Fig. 4(c). This confirms that the toroidal moment originates solely from the combination of chiral distortion and spin within a single MO$_6$ cluster[27]. Since the ME monopole, i.e., the antiferromagnetic domain, is invariant under this operation, the reversal of the magnetic toroidal domain originates from the reversal of the ferroaxial domain. This indicates that the magnetic toroidal domains are coupled with crystallographic ferroaxial domains, distinct from the ME monopole domains. Additionally, as shown in Fig. 4(d), reversing the ME monopole domains resulted in the same magnetic toroidal domains as in Fig. 4(a). Considering the point group 6/*m* symmetry and comparing the results in Figs. 4(a) and (c), we conclude that we successfully reversed the magnetic toroidal domains by reversing the ferroaxial domains without changing the direction of the ME monopole domains, which corresponds to the antiferromagnetic domains.

The successful observation and control of magnetic toroidal domains reveals the intricate coupling mechanism among different multipole orders in PbMn$_2$Ni$_6$Te$_3$O$_{18}$. Three multipole orders—magnetic toroidal, ME monopole, and ferroaxial—are coupled, because there are four types of domains: two types of antiferromagnetic domains and two types of crystal ferroaxial domains. There must be a constraint such as $T_z A_z M_0 > 0$ among these multipole orders. In Fig. 3(d) and previous studies[18,29], since the ferroaxial moment $A_z$ of the crystal is in a single domain state, aligning the ME monopole domains simultaneously aligns the magnetic toroidal domains, resulting in the visualization of ME monopole domains via magnetochiral imaging of magnetic toroidal domain. On the other hand, by aligning the ME monopole domain $M_0$ to be positive or negative, we expect that magnetic toroidal domains and ferroaxial domains will form completely identical domain structures, as shown in Figs. 4(e) and (f). In other words, ferroaxial domains can be indirectly observed by imaging magnetic toroidal domains.

In summary, by utilizing the coupling between multipole orders constrained by crystal structure, one can observe multipole domains that are typically difficult to detect. The magnetic toroidal domain imaging using directional dichroism is rapid and requires minimal equipment, as it only requires capturing two images—one in the low-temperature phase and one in the high-temperature phase—using unpolarized near-infrared light. Similar multipole couplings can be expected in various materials belonging to similar crystal and magnetic symmetries including MnTiO$_3$ and CoTe$_6$O$_{13}$. Such an imaging of multipole domains using coupling among different types of multipoles offers new opportunities for understanding and manipulating physical responses of the materials.

**Acknowledgements**

The authors thank M. Gen for fruitful discussion. This work was supported by JSPS KAKENHI (Grants No. 23H01120), JST FOREST (Grants No. JPMJFR2362).

**AUTHOR DECLARATIONS**

**Competing interests**

The authors have no conflicts to disclose.

**Author contributions**

**S. Aoyagi:** Data curation (lead); Formal analysis (lead); Investigation (lead); Methodology (equal); Project Administration (equal); Visualization (lead); Writing – original draft (lead); Writing – review & editing (equal). **S. Kitou:** Investigation (equal); Methodology (equal); Writing – review & editing (equal). **T. Arima:** Conceptualization (equal); Funding Acquisition (lead); Methodology (equal); Resources (lead); Supervision (lead); Writing – review & editing (equal); **Y. Tokunaga:** Conceptualization (lead); Formal Analysis (equal); Investigation (equal); Methodology (lead); Project Administration (equal); Resources (equal); Supervision (equal); Writing – original draft (equal); Writing – review & editing (equal).

**Data availability**

The data that support the findings of this study are available from the corresponding author upon reasonable request.